\documentclass[journal]{IEEEtran}
\usepackage{amsmath,amssymb,amsthm}
\usepackage{cite}
\usepackage[hidelinks]{hyperref}
\allowdisplaybreaks[2]
\newtheorem{theorem}{Theorem}
\newtheorem{lemma}{Lemma}
\newtheorem{corollary}{Corollary}
\theoremstyle{definition}
\newtheorem{definition}{Definition}
\newcommand{\X}{\mathcal X}
\newcommand{\Y}{\mathcal Y}
\newcommand{\Pcal}{\mathcal P}
\newcommand{\eps}{\varepsilon}
\newcommand{\TV}{\operatorname{TV}}
\newcommand{\Csum}{C_{\mathrm{sum}}}
\newcommand{\smin}{\sigma_{\min}}
\newcommand{\KL}{D_{\mathrm{KL}}}
\newcommand{\dd}{\mathrm d}
\newcommand{\TC}{\operatorname{TC}}
\begin{document}
\title{Local Stability of Treating Interference as Noise: An Explicit Sum-Capacity Guarantee}
\author{Xiqiang Qu, Luyao Fan, Jun Chen, Jia Wang, Zheng Dong}
\maketitle

\begin{abstract}
We study the local stability of treating interference as noise (TIN) for finite-alphabet interference channels with an arbitrary fixed number of users. A product reference channel is called regular if each component is strictly positive and has full row rank. We prove that every sufficiently small receiver-marginal perturbation of such a reference has sum capacity equal to the maximum TIN sum rate. The proof combines a multiuser genie-aided converse with a perturbation analysis that establishes its global tightness. The resulting explicit perturbation bounds provide a readily verifiable and physically interpretable sufficient criterion for the abstract information-theoretic notion of noisy interference. They show that noisy-interference channels arise naturally throughout neighborhoods of regular interference-free channels and rigorously justify independent coding and single-user decoding in these neighborhoods. 
\end{abstract}
\begin{IEEEkeywords}
Interference channel, treating interference as noise, sum capacity, genie-aided converse, less-noisy order, relative entropy, perturbation.
\end{IEEEkeywords}

\section{Introduction}\label{sec:intro}
\IEEEPARstart{T}{he} interference channel models communication between independent transmitter--receiver pairs sharing a common medium. Even for two users, its capacity region is not known in general. The Han--Kobayashi (HK) scheme \cite{HK} addresses interference through rate splitting: each message is divided into a part decoded at both receivers and a part decoded only at its intended receiver. Its equivalent simplified descriptions \cite{CMG} clarify the role of these two message layers. Treating interference as noise (TIN) is the special case in which both transmitters send only private information and each receiver decodes its intended message using the channel obtained by averaging over the interfering input.


For Gaussian interference channels, noisy-interference results establish the exact sum-capacity optimality of TIN under explicit conditions on the channel gains and power constraints \cite{SKC,AV,MK}. Genie-aided converses play a central role in these results. Specifically, the genie is designed to satisfy two complementary properties: usefulness, which ensures that the genie-aided sum-capacity admits a tractable characterization, and smartness, which ensures that the genie provides no rate improvement under the candidate sum-capacity-achieving input distribution.

For discrete memoryless interference channels, related lines of work identify regimes in which TIN is optimal, either within the HK achievable region or among all coding schemes. Liu, Nair, and Xia \cite{LNX} formulate a very-weak-interference condition under which TIN attains the maximum HK sum rate. They also derive a genie-aided sum-rate outer bound involving upper concave envelopes. Farsani \cite{Farsani} establishes a useful-genie converse for the two-user discrete memoryless interference channel. The associated noisy-interference criterion combines conditional less-noisy inequalities, required to hold for every product input distribution and auxiliary random variable, with smartness at an optimizer of the genie-aided objective. 

Subsequent work develops both the outer-bound method and the limitations of single-letter inner bounds. Liu, Nair, and Xia \cite{LNXenh} enlarge the genie construction to obtain an enhanced sum-rate outer bound; Liu's dissertation \cite{LiuThesis} gives a detailed treatment of genie-based bounds for discrete and Gaussian interference channels. Nair, Xia, and Yazdanpanah \cite{HKsub} show that two-letter coding can strictly outperform the HK region for clean Z-interference channels. 
Farsani and Khandani \cite{FK} extend the use of sequential Csisz\'ar--K\"orner identities to sum-rate converses and capacity results for multi-user interference channels.

Gohari and Nair \cite{GN} introduced the auxiliary-receiver approach, in which the additional receivers need not be enhanced versions of the original ones. Gohari, Nair, and Zhao \cite{GNZ24} derive further interference-channel outer bounds and characterize the capacity regions of two classes of discrete memoryless interference channels. Their more recent manuscript \cite{GNZ26} combines genie-aided and auxiliary-receiver add-and-subtract arguments in a unified outer bound for the two-user interference channel. These developments provide additional tools for capacity-region questions. 

Although the information-theoretic frameworks discussed above
provide general tools for analyzing discrete memoryless interference
channels, they do not generally yield explicit, readily checkable
conditions for optimality in terms of the physical channel transition
probabilities. Verifying their information-theoretic conditions
typically requires additional analysis tailored to the channel,
and tightness of the resulting bounds is therefore usually
established for particular channel classes, such as a subclass
of binary skewed-Z interference channels \cite{LNX}.


In this paper, we establish the sum-capacity optimality of TIN throughout explicit neighborhoods of interference-free channels. We call a product reference channel regular if each component is strictly positive and has full row rank. For every such $K$-user reference channel, we prove that all sufficiently small receiver-marginal perturbations have sum capacity equal to the maximum TIN sum rate. The proof combines a multiuser genie-aided converse with a perturbation analysis that establishes its global tightness.

The resulting perturbation bounds provide a readily verifiable and physically interpretable sufficient criterion for noisy interference, expressed directly in terms of the channel transition probabilities. The guarantee requires no symmetry or restriction on the interference topology, and a corollary gives a simpler closed-form bound on the allowable perturbation size. These results show that noisy-interference channels arise naturally throughout neighborhoods of regular interference-free channels, rigorously justifying independent coding and single-user decoding even as the maximizing input distributions and sum-capacity value vary within these neighborhoods.



The remainder of this paper is organized as follows. Section~\ref{sec:model} introduces the model and states the main result, which is proved in Section~\ref{sec:proof}. Section~\ref{sec:conclusion} concludes the paper.

\emph{Notation:} All alphabets are finite. For a positive integer $K$, write $[K]=\{1,\ldots,K\}$. For a tuple $x=(x_1,\ldots,x_K)$, let $x_{-i}=(x_j)_{j\ne i}$ denote the tuple obtained by omitting its $i$th coordinate. Probability distributions are represented as row vectors, and channel matrices have rows indexed by input symbols and columns indexed by output symbols. Thus, if $p$ is an input distribution and $A$ is a channel matrix, the corresponding output distribution is $pA$. The probability simplex on an alphabet $\mathcal A$ is denoted by $\Pcal(\mathcal A)$. For distributions $P,Q\in\Pcal(\mathcal A)$, their total variation distance is
$\TV(P,Q)=\frac12\|P-Q\|_1$.
For a tuple of input distributions $p=(p_i)_{i=1}^K$, the notation $p_{-i}$ in a channel average denotes the product distribution $\prod_{j\ne i}p_j$ over all inputs other than the $i$th.

The vector norms $\|\cdot\|_1$ and $\|\cdot\|_2$ denote the sum of absolute values and the Euclidean norm, respectively. The Euclidean norm of a tuple of vectors is the norm of the vector obtained by stacking its components. The matrix norms $\|\cdot\|_{\mathrm{op}}$ and $\|\cdot\|_{\mathrm F}$ denote the spectral and Frobenius norms, respectively. Entropies and mutual informations are measured in bits, whereas the Kullback--Leibler divergence $\KL$ is defined using natural logarithms. For an input distribution $p$ and a channel $A$, $I(p;A)$ denotes the mutual information induced by the joint distribution $P_{XY}(x,y)=p(x)A(y\mid x)$. For a real number $z$, write $[z]_+=\max\{z,0\}$. Finally, for a full-row-rank matrix $A$, define its smallest row singular value by
$\smin(A)=\sqrt{\lambda_{\min}(AA^{\mathsf T})}$,
where $A^{\mathsf T}$ denotes the transpose of $A$ and $\lambda_{\min}$ denotes the smallest eigenvalue.


\section{Problem Definition and Main Results}\label{sec:model}
\subsection{Channel Model and TIN Sum Rate}
\subsubsection{Discrete memoryless interference channel}
Consider a fixed integer $K\ge2$ and a $K$-user discrete memoryless
interference channel. Transmitter $i$ communicates an independent message
to receiver $i$, which observes only its own channel output. The finite
input and output alphabets are $\X_i$ and $\Y_i$, respectively, and the
single-use transition law is
\begin{equation}\label{eq:channel}
 W(y_1,\ldots,y_K\mid x_1,\ldots,x_K),
 \qquad x_i\in\X_i,\quad y_i\in\Y_i.
\end{equation}
Writing $x=(x_1,\ldots,x_K)$ and $y=(y_1,\ldots,y_K)$, memorylessness
means that, for any blocklength $N$,
\begin{equation}\label{eq:memorylesschannel}
 \begin{split}
 &W^{\otimes N}(y_1^N,\ldots,y_K^N\mid x_1^N,\ldots,x_K^N)\\
 &\qquad=\prod_{t=1}^N W(y_{1,t},\ldots,y_{K,t}\mid
                                     x_{1,t},\ldots,x_{K,t}).
 \end{split}
\end{equation}
The channel is fixed and known to every terminal. There is no feedback,
cooperation between transmitters or receivers, or input cost constraint.
The receiver-$i$ marginal channel is
\begin{equation}\label{eq:receivermarginals}
 W_i(y_i\mid x)=\sum_{y_{-i}}W(y_i,y_{-i}\mid x).
\end{equation}
We also write $W_i(y_i\mid x_i,x_{-i})$ to emphasize the intended input.
These marginals, rather than the joint dependence of the receiver outputs,
are the objects controlled by our perturbation assumption.

\subsubsection{Codes, achievable rates, and sum capacity}
 An $N$-use code
has message sets $[M_{i,N}]$, $i\in[K]$. The messages are independent and
uniform, so for every message tuple $m=(m_1,\ldots,m_K)$,
\begin{equation}\label{eq:messagedistribution}
 \Pr\{M_1=m_1,\ldots,M_K=m_K\}
   =\prod_{i=1}^K\frac1{M_{i,N}}.
\end{equation}
Each user has its own deterministic encoder and decoder,
\begin{align}
 f_{i,N}&:[M_{i,N}]\longrightarrow\X_i^N,
   \label{eq:encoderdefinition}\\
 g_{i,N}&:\Y_i^N\longrightarrow[M_{i,N}],
   \qquad i\in[K].\label{eq:decoderdefinition}
\end{align}
The transmitted codeword and decoded message are
\begin{equation}\label{eq:encodingdecoding}
 X_i^N=f_{i,N}(M_i),\qquad
 \widehat M_i=g_{i,N}(Y_i^N).
\end{equation}
The $K$ input sequences are independent because the messages are
independent and the encoders are separate. 

The average decoding error probability at receiver $i$ is
\begin{equation}\label{eq:individualerror}
 P_{e,i}^{(N)}=\Pr\{\widehat M_i\ne M_i\},
\end{equation}
where the average is over all messages and channel outputs. A
nonnegative rate tuple $(R_i)_{i=1}^K$ is achievable if there exists a
sequence of codes for which
\begin{equation}\label{eq:achieve}
 \liminf_{N\to\infty}\frac{\log_2 M_{i,N}}{N}\ge R_i,
 \qquad P_{e,i}^{(N)}\longrightarrow0\quad(i\in[K]).
\end{equation}
Let $\mathcal C(W)$ be the closure of the achievable rate tuples and
set
\begin{equation}\label{eq:sumcap}
 \Csum(W)=\sup_{(R_i)\in\mathcal C(W)}\sum_{i=1}^K R_i.
\end{equation}

The capacity region depends only on $W_1,\ldots,W_K$, not on the joint
coupling of the receiver outputs. This principle, also used for two
users in \cite{CMG}, lets us choose a convenient coupling of actual and
virtual receiver outputs in the genie construction without changing the
communication problem.

\subsubsection{The single-letter TIN objective}
Let
\begin{equation}\label{eq:productinputdomain}
 \mathcal D=\prod_{i=1}^K\Pcal(\X_i)
\end{equation}
be the compact set of product input distributions. A tuple
$p=(p_1,\ldots,p_K)\in\mathcal D$ induces
\begin{equation}\label{eq:productinputlaw}
 P_{X_1,\ldots,X_K}(x)=\prod_{i=1}^Kp_i(x_i)
\end{equation}
and the joint single-letter law
\begin{equation}\label{eq:TINjointlaw}
 P_p(x,y)=\left(\prod_{i=1}^Kp_i(x_i)\right)W(y\mid x).
\end{equation}
Unless otherwise stated, single-letter mutual informations associated
with $p$ are evaluated under this law. When receiver $i$ treats all
interfering inputs as noise, averaging over $p_{-i}$ gives the channel
\begin{equation}\label{eq:TINaveragedchannel}
 A_i^{p_{-i}}(y_i\mid x_i)
 =\sum_{x_{-i}}\left(\prod_{j\ne i}p_j(x_j)\right)
 W_i(y_i\mid x_i,x_{-i}).
\end{equation}
Its input--output joint law is
\begin{equation}\label{eq:TINpairlaw}
 P_{X_i,Y_i}(x_i,y_i)=p_i(x_i)A_i^{p_{-i}}(y_i\mid x_i).
\end{equation}
The TIN sum-rate objective is
\begin{equation}\label{eq:F}
 F_W(p)=\sum_{i=1}^K I_p(X_i;Y_i)
       =\sum_{i=1}^K I(p_i;A_i^{p_{-i}}).
\end{equation}
Since $F_W$ is continuous on the compact set $\mathcal D$, it attains its maximum. For each $p\in\mathcal D$, the single-user random-coding theorem \cite{Shannon48}, applied to the effective channels $A_i^{p_{-i}}$, shows that the sum rate $F_W(p)$ is achievable by independent coding and single-user decoding. Consequently,
\begin{equation}\label{eq:TINlowerbound}
 \Csum(W)\ge\max_{p\in\mathcal D}F_W(p).
\end{equation}

\subsection{Regular Product Channels}\label{sub:regular}
Our reference channel is the product of $K$ noninterfering
point-to-point links:
\begin{equation}\label{eq:product}
 V(y_1,\ldots,y_K\mid x_1,\ldots,x_K)
     =\prod_{i=1}^K V_i(y_i\mid x_i).
\end{equation}
In particular, receiver $i$'s output depends only on $x_i$. The
conditional independence of receiver outputs in \eqref{eq:product} is
a convenient representative of their joint coupling; the essential
reference property is that the receiver marginals are $V_i$.
Let $n_i=|\X_i|$ and $m_i=|\Y_i|$. We identify $V_i$ with an
$n_i\times m_i$ row-stochastic matrix. If $p_i$ is an input
probability row vector, then $p_iV_i$ is the output distribution.
\begin{definition}[Regular product channel]\label{def:regular}
The product reference \eqref{eq:product} is \emph{regular} when,
for every $i\in[K]$,
\begin{equation}\label{eq:regularconditions}
 V_i(y_i\mid x_i)>0\quad\text{for all }x_i,y_i,
 \qquad\operatorname{rank}(V_i)=n_i.
\end{equation}
We assume $n_i\ge2$; full row rank then requires $m_i\ge n_i$.
\end{definition}

\subsubsection{Strict positivity}
Strict positivity makes every reference transition probability
nonzero. Since there are finitely many entries and finitely many
users, their minimum $\alpha$ defined below is positive. Rowwise
perturbations with sufficiently small total variation therefore
preserve positivity of every receiver-marginal row and also make the
additional branch of the erasure genie a valid stochastic channel.

\subsubsection{Full row rank and distinguishability of input distributions}
Full row rank means that the linear map $v\mapsto vV_i$ has trivial
kernel on the space of real row vectors of length $n_i$.
Therefore
\begin{equation}\label{eq:inputinjectivity}
 p_iV_i=q_iV_i\quad\Longrightarrow\quad p_i=q_i
\end{equation}
for any two input distributions. The converse also holds for a
stochastic matrix. Indeed, if a nonzero row vector $v$ satisfies
$vV_i=0$, stochasticity implies
\begin{equation}\label{eq:kernelzerosum}
 v\mathbf1=vV_i\mathbf1=0.
\end{equation}
Let $\bar p_i$ be the uniform input distribution. For all sufficiently
small $h>0$, $\bar p_i\pm hv$ are two distinct probability
distributions, yet their output distributions are identical. Thus
injectivity on probability distributions is equivalent to full row
rank for a stochastic matrix.


The regularity assumptions hold for standard practical channel models. 
For example, a finite-constellation additive Gaussian noise channel 
followed by output quantization has strictly positive transition 
probabilities whenever the noise variance is positive and every 
quantization cell has positive Lebesgue measure. Full row rank also 
holds for common modulation and quantization schemes. In particular, binary antipodal signaling over a scalar additive
zero-mean Gaussian noise channel with positive noise variance,
followed by zero-threshold hard-decision detection, induces
a binary symmetric channel with crossover probability
$\theta\in(0,1/2)$. Its transition matrix has determinant
$1-2\theta>0$, so any product of such channels is regular.

\subsubsection{Reference parameters}
For each reference link, define
\begin{equation}\label{eq:rowsingularvalue}
 \sigma_i=\smin(V_i)
 =\min_{\|v\|_2=1}\|vV_i\|_2
 =\sqrt{\lambda_{\min}(V_iV_i^{\mathsf T})}>0.
\end{equation} 
We retain the individual parameters $n_i$ and $\sigma_i$
to capture differences among the reference links. We also
define the following parameters, including the uniform
bounds $n$ and $\sigma$ used in the closed-form corollary:
\begin{align}
 &\alpha=\min_{i,x_i,y_i}V_i(y_i\mid x_i),\qquad
 \sigma=\min_i\sigma_i,\label{eq:parameters}\\
 &n=\max_i n_i,\qquad q=K-1,\qquad N_i=\prod_{j\ne i}n_j.
 \label{eq:dimensions}
\end{align}
Here $N_i$ is the size of the \emph{joint} interfering-input alphabet
for receiver $i$, whereas $n_i$ is its intended-input alphabet size.
Since every link has at least two output symbols and all entries are
positive, $0<\alpha\le1/2$. Also
$\sigma_i^2\le\operatorname{tr}(V_iV_i^{\mathsf T})/n_i\le1$, so
$0<\sigma_i\le1$.
The symmetric matrix
\begin{equation}\label{eq:interactionmatrix}
 M_{ij}=\begin{cases}
   \sqrt{n_in_j},&i\ne j,\\0,&i=j,
 \end{cases}
 \qquad\rho=\lambda_{\max}(M)
\end{equation}
records the sizes of pairwise mixed-input terms in the tightness
estimate. For $K=2$, $\rho=\sqrt{n_1n_2}$; when every $n_i=n$,
$\rho=n(K-1)$. 

\subsubsection{Measuring perturbations of the receiver marginals}
Let $W$ be any $K$-user channel with the same alphabets as $V$, and
measure its distance to the reference by
\begin{equation}\label{eq:distance}
 d(W,V)=\max_{i,x}\frac12\sum_{y_i}
   |W_i(y_i\mid x)-V_i(y_i\mid x_i)|.
\end{equation}
Thus $d(W,V)\le\eps$ requires \emph{every} receiver-marginal
conditional distribution, at \emph{every} input tuple, to lie within
total variation $\eps$ of its reference row. The maximum is uniform
over inputs and does not depend on any candidate input distribution.
It remains valid when we optimize over $\mathcal D$, including its
boundary.

Two rowwise consequences of this assumption are
\begin{align}
 \|W_i(\cdot\mid x)-V_i(\cdot\mid x_i)\|_1
   &\le2\eps,\label{eq:rowLoneperturbation}\\
 |W_i(y_i\mid x)-V_i(y_i\mid x_i)|
   &\le\eps.\label{eq:coordinateperturbation}
\end{align}
The first follows directly from total variation. The second uses
the fact that the positive and negative parts of the difference of
two probability vectors have equal total mass, so no coordinate
change exceeds the total variation distance. Consequently
\begin{equation}\label{eq:perturbedentrylowerbound}
 W_i(y_i\mid x)\ge\alpha-\eps.
\end{equation}
In particular, every receiver-marginal entry stays positive when
$\eps<\alpha$.

The same bounds hold after averaging over arbitrary product inputs
on the other $K-1$ users. For example, by convexity of the
$\ell_1$ norm and \eqref{eq:TINaveragedchannel},
\begin{align}\label{eq:averagedTVperturbation}
 &\TV\bigl(A_i^{p_{-i}}(\cdot\mid x_i),
                         V_i(\cdot\mid x_i)\bigr)\nonumber\\
 &\quad\le\sum_{x_{-i}}\left(\prod_{j\ne i}p_j(x_j)\right)
        \TV\bigl(W_i(\cdot\mid x_i,x_{-i}),
                         V_i(\cdot\mid x_i)\bigr)\nonumber\\
 &\quad\le\eps.
\end{align}
This estimate also holds if the interfering-input averaging law is
correlated, because the proof needs only nonnegative weights summing
to one. In particular, all entries of any such averaged channel are
at least $\alpha-\eps$.

The perturbations may alter the intended links and introduce arbitrary
dependence on any subset of interferers; they need not respect a given
network topology, additivity, or symmetry. Finally, the distance
\eqref{eq:distance} is a metric on collections of receiver-marginal
channels but only a pseudometric on full joint transition laws:
$d(W,V)=0$ allows different output couplings. 
This is exactly the freedom relevant
to capacity. In particular, the permitted perturbations do not have
to preserve the product coupling in \eqref{eq:product}. 

\subsection{A Multiuser Useful-Genie Converse}\label{sub:farsani-criterion}
A channel $A$ is less noisy than a channel $B$ with the same input alphabet if
$I(U;Z_A)\ge I(U;Z_B)$ for every joint distribution of the input and an auxiliary random variable $U$, where $Z_A$ and $Z_B$ are generated through the respective channels and are conditionally independent of $U$ given the input \cite{KMcompare}. This comparison depends only on the two conditional output distributions. We use the following multiuser extension of the useful-genie principle of \cite{Farsani}. Its proof is given in Appendix \ref{app:lemma1}. 

\begin{lemma}[Multiuser useful-genie converse]\label{lem:farsani}
Fix a memoryless augmentation with outputs $Z_i=(Y_i,T_i)$ that preserves every receiver marginal and satisfies
\begin{equation}\label{eq:virtualproduct}
 P(t_1,\ldots,t_K\mid x)=\prod_{i=1}^K K_i(t_i\mid x_i).
\end{equation}
Choose $\eta_i\in[0,1]$ such that
\begin{equation}\label{eq:budget}
 \sum_{i\ne j}\eta_i\le1,\qquad j\in[K].
\end{equation}
For each $i\in[K]$, let
$\mathcal X_{-i}=\prod_{j\ne i}\mathcal X_j$ and
$\mathcal T_{-i}=\prod_{j\ne i}\mathcal T_j$.
Let $\mathsf e_i$ be an erasure symbol not in $\mathcal T_{-i}$.
Define $E_i$ as the channel from $\mathcal X_{-i}$ to
$\mathcal T_{-i}\cup\{\mathsf e_i\}$ with transition probabilities
\begin{equation}\label{eq:erasure-channel}
 E_i(\widetilde t_i\mid x_{-i})=
 \begin{cases}
  \displaystyle \eta_i\prod_{j\ne i}K_j(t_j\mid x_j),
       & \widetilde t_i=t_{-i}\in\mathcal T_{-i},\\[1ex]
  1-\eta_i, & \widetilde t_i=\mathsf e_i.
 \end{cases}
\end{equation}
Thus, $E_i$ generates $T_{-i}$ through the product channel
$\bigotimes_{j\ne i}K_j$ and reveals the entire vector with probability $\eta_i$, otherwise emitting $\mathsf e_i$. The erasure event is independent of the inputs and all other channel randomness. Let $\widetilde T_i$ denote the output of $E_i$.

For each fixed $x_i\in\mathcal X_i$, define $L_{i,x_i}$ as the channel from $\mathcal X_{-i}$ to $\mathcal Y_i\times\mathcal T_i$ with transition probabilities
\begin{equation}\label{eq:fixed-input-channel}
 L_{i,x_i}(y_i,t_i\mid x_{-i})
 =
 P_{Y_i,T_i\mid X}
 (y_i,t_i\mid x),
 \qquad x=(x_i,x_{-i}),
\end{equation}
where the coordinates of $x$ retain their original ordering and the conditional distribution on the right is specified by the fixed augmentation. Thus, $L_{i,x_i}$ describes the augmented observation $Z_i=(Y_i,T_i)$ when the intended input is held fixed at $x_i$ and the remaining inputs vary.

Suppose that
\begin{equation}\label{eq:farsani-conditions}
 E_i \text{ is less noisy than } L_{i,x_i},
 \qquad i\in[K],\quad x_i\in\mathcal X_i.
\end{equation}
Then
\begin{equation}\label{eq:geniebound}
 \Csum(W)\le\max_{p\in\mathcal D}J(p),\qquad
 J(p)=\sum_{i=1}^K I_p(X_i;Y_i,T_i).
\end{equation}
\end{lemma}

Lemma~\ref{lem:farsani} can be interpreted as providing receiver $i$
with genie information $T_i$ in addition to its original observation
$Y_i$. Providing this information cannot reduce the achievable rates.
The less-noisy comparisons in \eqref{eq:farsani-conditions},
together with the allocation constraints in \eqref{eq:budget},
yield a converse matching the achievable sum rate
$\max_{p\in\mathcal D}J(p)$ of the augmented channel.
This is the usefulness property of the genie.
The channels $E_i$ serve only as comparison channels in the
converse; their outputs are not additional observations supplied to
receiver $i$. They quantify how much information about the interfering
inputs can be conveyed by the augmented observation when the intended
input is fixed.

To establish TIN optimality for the original channel, we must also
make this upper bound tight. We construct the genie using a maximizer
$p^*$ of $F_W$ so that
$I_{p^*}(X_i;T_i\mid Y_i)=0$ for every $i$.
Thus, the genie provides no rate improvement at $p^*$, giving
$J(p^*)=F_W(p^*)$; this is the smartness property.
Smartness alone does not rule out a larger value of $J$ at another
input distribution. The subsequent perturbation analysis establishes
the global inequality $J(p)\le F_W(p^*)$ for all
$p\in\mathcal D$, thereby matching the converse with the achievable
TIN sum rate.

\subsection{Explicit Sum-Capacity Guarantee}\label{sub:results}
Our main contribution is an explicit sufficient condition under which
the useful-genie bound in Lemma~\ref{lem:farsani} coincides with the
maximum TIN sum rate. The condition is expressed in terms of the
parameters of a regular product reference channel and the size of
the receiver-marginal perturbation. It yields a positive neighborhood
of every such reference throughout which independent coding and
single-user decoding achieve sum capacity.

For $\eps\ge0$ and $0<\tau<1$, define
\begin{align}
 &a=\alpha-\eps,\qquad b=\alpha-\frac{1+\tau}{1-\tau}\eps,\label{eq:ab}\\
 &s_i=\sigma_i-\sqrt{2n_i}\eps,\qquad S_i=\prod_{j\ne i}s_j.\label{eq:s}
\end{align}
When $b>0$ and all $s_i>0$, set
\begin{align}
 &u_i=\frac{N_i\eps^2}{(1-\tau)b\tau^q S_i^2},\label{eq:ui}\\
 &\ell_j=\sum_{i\ne j}u_i,\qquad c_j=\tau-\tau^q\ell_j.\label{eq:loads}
\end{align}
For a nonnegative vector $d\in\mathbb R^K$, define the following function, which is positively homogeneous of degree two:
\begin{align}\label{eq:Psi}
 \Psi_{\eps,\tau}(d)
 &=\sum_i c_i s_i^2d_i^2\nonumber\\
 &\quad+(1-\tau)\sum_i\left[s_i d_i-
 \frac{\sqrt2\eps}{1-\tau}\sum_{j\ne i}\sqrt{n_j}d_j\right]_+^2\nonumber\\
 &\quad-K\eps\ln\frac1a\,d^{\mathsf T}Md,
\end{align}
and its minimum on the nonnegative unit sphere,
\begin{equation}\label{eq:Gamma}
\Gamma(\eps,\tau)=\min_{d\ge0:\,\|d\|_2=1}\Psi_{\eps,\tau}(d).
\end{equation}

\begin{theorem}[Local stability of multiuser TIN]\label{thm:main}
Let $V$ be a regular $K$-user product channel. Suppose $\eps\ge0$ and some $\tau\in(0,1)$ satisfy
\begin{align}
&b>0,\qquad s_i>0\quad(i\in[K]),\label{eq:basic}\\
 &\ell_j\le1\quad(j\in[K]),\label{eq:useful}\\
 &\Gamma(\eps,\tau)\ge0.\label{eq:tight}
\end{align}
Then every interference channel $W$ on the same alphabets with $d(W,V)\le\eps$ has sum capacity
\begin{equation}\label{eq:main}
 \Csum(W)=\max_{p\in\mathcal D}F_W(p).
\end{equation}
The inequalities in \eqref{eq:useful}--\eqref{eq:tight} and the distance bound are non-strict. If $\Gamma(\eps,\tau)>0$, the maximizing product input in \eqref{eq:main} is unique.
\end{theorem}

Condition~\eqref{eq:useful} ensures that the genie-aided converse
applies, while condition~\eqref{eq:tight} ensures that its bound
matches the maximum TIN sum rate. Together, they identify a
neighborhood in which independent coding and single-user decoding
remain sum-capacity optimal.

Optimizing over the genie parameter $\tau$ gives the largest
perturbation radius certified by our criterion:
\begin{equation}\label{eq:certifiedradius}
 \eps_{\mathrm{cert}}=\max\{\eps\ge0:\ \exists\tau\in(0,1)
 \text{ satisfying \eqref{eq:basic}--\eqref{eq:tight}}\}.
\end{equation}
Section~\ref{sub:radius} shows that this maximum is attained
and strictly positive, and that the guarantee includes the
boundary of the certified neighborhood.



The following more conservative formula avoids optimization entirely. Let
\begin{equation}\label{eq:explicitconstants}
 b_0=\frac\alpha4,\qquad L_0=\ln\frac4{3\alpha},\qquad
 C_K=(2K-1)\sqrt{2n},
\end{equation}
and define
\begin{align}
 &\eps_{\mathrm U}
 =\frac{\sigma^q}{q\sigma^{q-1}\sqrt{2n}
       +\sqrt{2^{q+1}q n^q/b_0}},\label{eq:radiusU}\\
 &\eps_{\mathrm G}
 =\frac{\sigma^2}{2\sigma C_K+2K\rho L_0},\label{eq:radiusG}\\
 &\eps_{\mathrm{cf}}=\min\left\{\frac\alpha4,
          \eps_{\mathrm U},\eps_{\mathrm G}\right\}.\label{eq:radius}
\end{align}
\begin{corollary}[Closed-form radius]\label{cor:explicit}
For every regular $K$-user product reference, $\eps_{\mathrm{cf}}>0$. Every $W$ satisfying $d(W,V)\le\eps_{\mathrm{cf}}$ satisfies \eqref{eq:main}. In particular, $\eps_{\mathrm{cert}}\ge\eps_{\mathrm{cf}}$.
\end{corollary}

\section{Proof of the Main Result}\label{sec:proof}
The proof proceeds by constructing a genie for which the outer bound
in Lemma~\ref{lem:farsani} matches the maximum TIN sum rate. We begin
with elementary norm and relative-entropy estimates that quantify
the effect of perturbing a regular product channel. We then choose
a TIN-maximizing input distribution and construct a genie that
provides no rate improvement at this input. For sufficiently small
perturbations, the estimates ensure that the genie satisfies the
usefulness conditions of Lemma~\ref{lem:farsani}. Finally, we show
that no other product input distribution yields a larger genie-aided
sum rate. This global comparison accounts for simultaneous changes
in several users' input distributions and establishes equality
between the outer bound and the achievable TIN sum rate.

\subsection{Zero-Sum and Relative-Entropy Estimates}\label{sub:estimates}
We first establish the norm and divergence inequalities used in
the genie usefulness comparison and the global tightness analysis.
All relative entropies in this subsection are measured in nats.

If a real vector $v$ has zero sum, write $v=v^+-v^-$, where
$v^+,v^-$ are nonnegative and have disjoint supports. Both parts
have $\ell_1$ norm $\|v\|_1/2$, so
\begin{equation}\label{eq:zerosum}
 \|v\|_2^2
 =\|v^+\|_2^2+\|v^-\|_2^2
 \le\|v^+\|_1^2+\|v^-\|_1^2
 =\frac12\|v\|_1^2.
\end{equation}
Applying \eqref{eq:zerosum} to the difference of two probability
vectors gives
\begin{equation}\label{eq:tv2}
 \|P-Q\|_2\le\sqrt2\,\TV(P,Q).
\end{equation}

We will also use the divergence bounds
\begin{align}
 &\KL(P\|Q)\ge\tfrac12\|P-Q\|_1^2
                 \ge\|P-Q\|_2^2,\label{eq:kl-lower}\\
 &\KL(P\|Q)\le\frac{\|P-Q\|_2^2}{2b}
 \quad\text{if }P_y,Q_y\ge b>0\text{ for every }y.
 \label{eq:kl-upper}
\end{align}
The first follows from Pinsker's inequality
\cite{Csiszar67,Kullback67} and \eqref{eq:zerosum}.
To prove the second, put $v=P-Q$ and define
$g(t)=\KL(Q+tv\|Q)$ for $0\le t\le1$. Direct differentiation yields
\begin{equation}\label{eq:kl-derivatives}
 g(0)=g'(0)=0,\qquad
 g''(t)=\sum_y\frac{v_y^2}{Q_y+tv_y}.
\end{equation}
Integrating twice gives
\begin{equation}\label{eq:kl-integral}
 \KL(P\|Q)=\int_0^1(1-t)\sum_y
       \frac{v_y^2}{Q_y+tv_y}\,\dd t.
\end{equation}
Since $Q_y+tv_y=(1-t)Q_y+tP_y\ge b$, the integral is at most
$\|v\|_2^2/(2b)$, proving \eqref{eq:kl-upper}.
See \cite{SasonVerdu} for related divergence comparisons.

The definition of $\sigma_i$ gives
\begin{equation}\label{eq:referencequantitativeinjectivity}
 \|vV_i\|_2\ge\sigma_i\|v\|_2
\end{equation}
for every real row vector $v$. Combining this with
\eqref{eq:kl-lower}, we obtain
\begin{equation}\label{eq:referenceKLseparation}
 \KL(p_iV_i\|q_iV_i)
 \ge\|(p_i-q_i)V_i\|_2^2
 \ge\sigma_i^2\|p_i-q_i\|_2^2
\end{equation}
for any input distributions $p_i,q_i$.

We next derive two uniform consequences of the perturbation
assumption. First, fix receiver $i$ and let $A_i$ be obtained by
averaging $W_i(\cdot\mid x_i,x_{-i})$ over any interfering-input
distribution, which need not be a product distribution.
By \eqref{eq:averagedTVperturbation}, each row of $A_i$ is within
total variation $\eps$ of the corresponding row of $V_i$.
Hence \eqref{eq:tv2} gives
\begin{align}\label{eq:matrixperturb}
 \|A_i-V_i\|_{\mathrm{op}}
 &\le\|A_i-V_i\|_{\mathrm F}\nonumber\\
 &\le\sqrt{n_i}\max_{x_i}
    \|(A_i-V_i)(\cdot\mid x_i)\|_2\nonumber\\
 &\le\sqrt{2n_i}\,\eps.
\end{align}
For every real row vector $v$ on $\X_i$, the reverse triangle
inequality and \eqref{eq:referencequantitativeinjectivity} imply
\begin{align}\label{eq:singular}
 \|vA_i\|_2
 &\ge\|vV_i\|_2-\|v(A_i-V_i)\|_2\nonumber\\
 &\ge(\sigma_i-\sqrt{2n_i}\,\eps)\|v\|_2
 =s_i\|v\|_2.
\end{align}
Thus, when $s_i>0$, every such averaged channel has full row rank.
The estimate holds for every real row vector of length $n_i$,
which will allow us to apply it to tensor products in the
usefulness comparison.


Second, fix $x_i$ and let $v$ be a real row vector on
$\X_{-i}=\prod_{j\ne i}\X_j$ satisfying
$\sum_{x_{-i}}v(x_{-i})=0$. Define
\begin{equation}\label{eq:ddef}
 d_{i,x_i}(y)=\sum_{x_{-i}}v(x_{-i})
                  W_i(y\mid x_i,x_{-i}).
\end{equation}
Because the reference row does not depend on $x_{-i}$, the
zero-sum condition gives
\begin{equation}\label{eq:referencecancellation}
 d_{i,x_i}(y)
 =\sum_{x_{-i}}v(x_{-i})
   \bigl[W_i(y\mid x_i,x_{-i})-V_i(y\mid x_i)\bigr].
\end{equation}
Applying the triangle inequality and
\eqref{eq:rowLoneperturbation} yields
\begin{align}\label{eq:d1}
 \|d_{i,x_i}\|_1
 &\le\sum_{x_{-i}}|v(x_{-i})|
   \|W_i(\cdot\mid x_i,x_{-i})-V_i(\cdot\mid x_i)\|_1
   \nonumber\\
 &\le2\eps\|v\|_1.
\end{align}
Similarly, \eqref{eq:tv2} and Cauchy--Schwarz give
\begin{align}\label{eq:dbounds}
 \|d_{i,x_i}\|_2
 &\le\sum_{x_{-i}}|v(x_{-i})|
   \|W_i(\cdot\mid x_i,x_{-i})-V_i(\cdot\mid x_i)\|_2
   \nonumber\\
 &\le\sqrt2\eps\|v\|_1
 \le\sqrt{2N_i}\eps\|v\|_2.
\end{align}

The cancellation in \eqref{eq:referencecancellation} also applies
when $v$ is a tensor product containing at least one zero-sum
factor, since the entire product then has zero sum. This includes
both the first-order and higher-order terms arising when several
interfering-input distributions vary; these terms will be bounded
in the global tightness analysis. 

\subsection{Erasure Genie and Usefulness}\label{sub:genie}
Fix parameters satisfying \eqref{eq:basic}--\eqref{eq:useful}. By compactness choose $p^*\in\arg\max_{p\in\mathcal D}F_W(p)$, and keep it fixed when constructing the genie. Define
\begin{align}
 &B_i=A_i^{p^*_{-i}},\label{eq:B}\\
 &G_i(y\mid x)=\frac{W_i(y\mid x)-\tau B_i(y\mid x_i)}{1-\tau}.
 \label{eq:G}
\end{align}
The rows of $G_i$ sum to one. Coordinatewise perturbation bounds give
\begin{equation}\label{eq:Gpositive}
 G_i(y\mid x)\ge V_i(y\mid x_i)-\frac{1+\tau}{1-\tau}\eps
 \ge b>0.
\end{equation}
Thus $G_i$ is a valid channel. Put $\mathcal T_i=\Y_i\cup\{\bot\}$, with $\bot$ a new symbol, and define the joint receiver--genie marginal
\begin{equation}\label{eq:L}
 L_i(y,t\mid x)=
 \begin{cases}
  \tau B_i(y\mid x_i),&t=y,\\
  (1-\tau)G_i(y\mid x),&t=\bot,\\
  0,&\text{otherwise}.
 \end{cases}
\end{equation}
Summing over $t$ gives $W_i$; summing over $y$ gives
\begin{equation}\label{eq:K}
 K_i(t\mid x_i)=\begin{cases}
  \tau B_i(t\mid x_i),&t\in\Y_i,\\
  1-\tau,&t=\bot.
 \end{cases}
\end{equation}
Choose the simultaneous augmentation $\widetilde W=\prod_i L_i$, conditionally on all inputs. It preserves every $W_i$ and satisfies \eqref{eq:virtualproduct}. Its different coupling of the actual outputs is immaterial by marginal invariance.

Fix $i,x_i$ and arbitrary distributions $P,Q$ on
$\mathcal X_{-i}$, including correlated distributions.
The branch $t\in\Y_i$ occurs with the same probability $\tau$
under both input laws. In that branch the joint output is $(y,y)$
with distribution $B_i(y\mid x_i)$, independent of $X_{-i}$,
and therefore contributes zero divergence. In the other branch
the output is $(y,\bot)$ with distribution induced by $G_i$.
The two branches have disjoint output supports, so their KL
divergences add with their common weights, giving
\begin{equation}\label{eq:ZKL}
 \KL(PL_{i,x_i}\|QL_{i,x_i})
 =(1-\tau)\KL(PG_i\|QG_i).
\end{equation}
Here and below $G_i$ has the fixed intended input $x_i$. Applying \eqref{eq:dbounds} to $v=P-Q$, the $B_i$ term cancels and gives
\begin{equation}\label{eq:Gdifference}
 \|PG_i-QG_i\|_2\le\frac{\sqrt{2N_i}\eps}{1-\tau}\|P-Q\|_2.
\end{equation}
Together with \eqref{eq:kl-upper} and \eqref{eq:Gpositive}, this proves
\begin{equation}\label{eq:GKL}
 \KL(PG_i\|QG_i)\le
 \frac{N_i\eps^2}{(1-\tau)^2b}\|P-Q\|_2^2.
\end{equation}
Set $\mathcal B_{-i}=\bigotimes_{j\ne i}B_j$ and $\mathcal K_{-i}=\bigotimes_{j\ne i}K_j$. For clarity, the Gram matrix of the product channel satisfies
\begin{equation}\label{eq:Gramproduct}
 \mathcal B_{-i}\mathcal B_{-i}^{\mathsf T}
 =\bigotimes_{j\ne i}(B_jB_j^{\mathsf T}).
\end{equation}
The eigenvalues of a Kronecker product are all products of
eigenvalues of its factors. Each factor is positive definite,
and \eqref{eq:singular} bounds its smallest eigenvalue below
by $s_j^2$. Thus the smallest row singular value of
$\mathcal B_{-i}$ is at least $S_i$. Consequently,
\eqref{eq:kl-lower} gives
\begin{equation}\label{eq:BKL}
 \KL(P\mathcal B_{-i}\|Q\mathcal B_{-i})\ge S_i^2\|P-Q\|_2^2.
\end{equation}
The output of $\mathcal K_{-i}$ reveals the retained-index set
$S\subseteq[K]\setminus\{i\}$. Because each virtual channel
retains independently with probability $\tau$, its pattern law
is $\Pr(S)=\tau^{|S|}(1-\tau)^{q-|S|}$, independent of the input.
For each $S\subseteq[K]\setminus\{i\}$, let $P_S$ and $Q_S$
denote the marginals of $P$ and $Q$ on
$\mathcal X_S=\prod_{j\in S}\mathcal X_j$, and let
$\mathcal B_S=\bigotimes_{j\in S}B_j$ be the channel from
$\mathcal X_S$ to $\prod_{j\in S}\mathcal Y_j$.
For $S=\varnothing$, interpret both output laws as the same
point mass, so the corresponding divergence is zero.
The chain rule for divergence on the disjoint pattern events
gives the exact decomposition
\begin{align}\label{eq:erasurepatternsum}
 &\KL(P\mathcal K_{-i}\|Q\mathcal K_{-i})\nonumber\\
 &\quad=\sum_{S\subseteq[K]\setminus\{i\}}
   \tau^{|S|}(1-\tau)^{q-|S|}
   \KL(P_S\mathcal B_S\|Q_S\mathcal B_S).
\end{align}
Every summand is nonnegative. Keeping only the event that
all $q$ outputs are retained gives
\begin{equation}\label{eq:producterasureKL}
 \KL(P\mathcal K_{-i}\|Q\mathcal K_{-i})
 \ge\tau^q\KL(P\mathcal B_{-i}\|Q\mathcal B_{-i}).
\end{equation}
Choose $\eta_i=u_i$ from \eqref{eq:ui}. Each $u_i$ is
nonnegative and appears in at least one load
$\ell_j=\sum_{k\ne j}u_k$ because $K\ge2$.
Thus \eqref{eq:useful} implies $u_i\le1$ and, for each
virtual-output index $j$,
$\sum_{i\ne j}\eta_i=\ell_j\le1$.
This verifies both $\eta_i\in[0,1]$ for every $i\in[K]$ and
the budget constraint \eqref{eq:budget} in
Lemma~\ref{lem:farsani}.
The construction uses two different erasures. Each $K_j$
retains its own $B_j$-output with probability $\tau$,
while the outer channel $E_i$ retains the entire vector
$T_{-i}$ with probability $u_i=\eta_i$. These decisions
are independent. Thus $E_i$ reveals all $q=K-1$ direct
outputs with probability $u_i\tau^q$, which is the
coefficient used in the next comparison.
Applying the independent outer erasure to $\mathcal K_{-i}$
makes $E_i$ reveal its output with probability $u_i$ and
otherwise emit a new symbol; its two output branches again have
disjoint supports. Therefore
\begin{align}\label{eq:EKL}
 \KL(PE_i\|QE_i)
 &=u_i\KL(P\mathcal K_{-i}\|Q\mathcal K_{-i})\nonumber\\
 &\ge\frac{N_i\eps^2}{(1-\tau)b}\|P-Q\|_2^2\nonumber\\
 &\ge\KL(PL_{i,x_i}\|QL_{i,x_i}).
\end{align}
This includes $\eps=0$, when both sides in the needed comparison are zero. Pairwise divergence dominance implies the less-noisy order: for any auxiliary $U$ and channel $A$,
\begin{equation}\label{eq:MI-KL}
 I(U;Z_A)=\frac1{\ln2}\sum_uP_U(u)
 \KL(P_{X|u}A\|P_XA).
\end{equation}
For any auxiliary law $P_{U,X_{-i}}$, apply
\eqref{eq:EKL} to $P=P_{X_{-i}|U=u}$ and
$Q=P_{X_{-i}}$, multiply by $P_U(u)$, and sum over $u$.
The identity \eqref{eq:MI-KL} gives
$I(U;\widetilde T_i)\ge I(U;Z_i)$, where
$\widetilde T_i$ and $Z_i$ are generated through $E_i$
and $L_{i,x_i}$, respectively.
This holds for arbitrary, possibly correlated, interfering inputs.
Since the KL comparison is uniform over the fixed intended
input $x_i$, it proves \eqref{eq:farsani-conditions}
for every $i,x_i$.
Hence the multiuser converse \eqref{eq:geniebound} applies.

We will also need the stronger comparison before the erasures. Combining \eqref{eq:GKL} and \eqref{eq:BKL} gives
\begin{equation}\label{eq:rowcomparison}
 \KL(PG_i\|QG_i)\le\kappa_i
 \KL(P\mathcal B_{-i}\|Q\mathcal B_{-i}).
\end{equation}
Here $\kappa_i=\tau^q u_i/(1-\tau)$. 

\subsection{Global Tightness from the Usefulness Comparison}\label{sub:tightness}
\subsubsection{Smartness and tangency}
Keep the genie fixed at $p^*$ and vary the product input $p$. Write
\begin{align}
 &C_i^p=\frac{A_i^{p_{-i}}-\tau B_i}{1-\tau},\qquad
 H_i^p=C_i^p-B_i,\label{eq:C}\\
 &\delta_i=p_i-p_i^*,\qquad\nu_i^*=p_i^*B_i.\label{eq:Hdelta}
\end{align}
For each $x_i\in\mathcal X_i$, write
$B_{i,x_i}=B_i(\cdot\mid x_i)$ and
$C_{i,x_i}^p=C_i^p(\cdot\mid x_i)$
for the corresponding channel rows.
The channel $C_i^p$ is the average of $G_i$ over $p_{-i}$, so it is stochastic with entries at least $b$. The branch indicator in \eqref{eq:L} is independent of $X_i$. On the retained branch the pair $(Y_i,T_i)$ is equivalent to the output of $B_i$, whereas on the erased branch it is the output of $C_i^p$. Thus
\begin{align}
 &F_i(p)=I(p_i;A_i^{p_{-i}}),\label{eq:Fcomponent}\\
 &J_i(p)=\tau I(p_i;B_i)+(1-\tau)I(p_i;C_i^p),
 \qquad J=\sum_iJ_i.\label{eq:Javerage}
\end{align}
At $p=p^*$, $A_i^{p^*_{-i}}=C_i^{p^*}=B_i$, and hence
\begin{equation}\label{eq:smart}
 J(p^*)=F_W(p^*).
\end{equation}
Equivalently, conditional on $Y_i$ the genie reveals that output with probability $\tau$ and otherwise erases it, independently of $X_i$. This proves $I_{p^*}(X_i;T_i\mid Y_i)=0$.

The first derivatives also agree. Let $p(t)=p^*+t\delta$,
$A_i(t)=A_i^{p_{-i}(t)}$, and
$C_i(t)=(A_i(t)-\tau B_i)/(1-\tau)$.
Primes denote differentiation with respect to $t$, applied
entrywise to vector- and matrix-valued functions.
At $t=0$, both channels equal $B_i$, and
$C_i'(0)=A_i'(0)/(1-\tau)$.
To differentiate both the input law and the channel, let
$\mu_i(t)=p_i(t)A_i(t)$ be the output distribution. In nats,
\begin{align}
 (\ln2)F_i(p(t))
 &=\sum_{x_i,y}p_i(t)(x_i)A_i(t)(y\mid x_i)
       \ln A_i(t)(y\mid x_i)\nonumber\\
 &\quad-\sum_y\mu_i(t)(y)\ln\mu_i(t)(y).
\end{align}
Terms from differentiating the logarithms alone cancel:
their sum is
$\sum_{x_i,y}p_i(t)(x_i) A'_i(t)(y\mid x_i)
-\sum_y\mu'_i(t)(y)=0$,
because channel rows and the output distribution
have total mass one.
Differentiating the remaining weights by the product rule gives
\begin{align}\label{eq:mutualderivative}
 &(\ln2)\frac{\dd}{\dd t}F_i(p(t))\nonumber\\
 &=\sum_{x_i,y}\delta_i(x_i)A_i(t)(y\mid x_i)
 \ln\frac{A_i(t)(y\mid x_i)}{\mu_i(t)(y)}\nonumber\\
 &\quad+\sum_{x_i,y}p_i(t)(x_i) A'_i(t)(y\mid x_i)
 \ln\frac{A_i(t)(y\mid x_i)}{\mu_i(t)(y)}.
\end{align}
At $t=0$, $A_i(0)=B_i$ and $\mu_i(0)=\nu_i^*$.
The $\delta_i$ contribution equals
$\sum_{x_i}\delta_i(x_i)\KL(B_{i,x_i}\|\nu_i^*)$.
Consequently, \eqref{eq:mutualderivative} becomes
\begin{align}\label{eq:Fdirection}
 &(\ln2)\left.\frac{\dd}{\dd t}F_i(p(t))\right|_{t=0}\nonumber\\
 &=\sum_{x_i}\delta_i(x_i)\KL(B_{i,x_i}\|\nu_i^*)\nonumber\\
 &\quad+\sum_{x_i,y}p_i^*(x_i) A'_i(0)(y\mid x_i)
       \ln\frac{B_i(y\mid x_i)}{\nu_i^*(y)}.
\end{align}
For $J_i$, the $B_i$ branch contributes $\tau$ times the
input-variation term, while the $C_i(t)$ branch contributes
$1-\tau$ times that term and has weighted channel variation
$(1-\tau) C'_i(0)= A'_i(0)$.
Thus its derivative is exactly \eqref{eq:Fdirection}. Writing
$DG(p^*)[\delta]
=\left.\frac{\dd}{\dd t}G(p^*+t\delta)\right|_{t=0^+}$
for the directional derivative of $G$, we conclude that
\begin{equation}\label{eq:tangent}
 DJ(p^*)[\delta]=DF_W(p^*)[\delta]\le0,
 \qquad \delta=p-p^*,\quad p\in\mathcal D.
\end{equation}
The last inequality is the one-sided first-order condition along the feasible segment from $p^*$ to $p$. Positivity of the channel and output probabilities makes these derivatives well defined even when $p^*$ is on the boundary. Smartness and tangency alone do not establish that $p^*$ maximizes $J$; the following global estimate does.

\begin{lemma}[Unified multiuser tightness estimate]\label{lem:tightness}
Under \eqref{eq:basic}--\eqref{eq:useful}, for every $p\in\mathcal D$, let $d_i=\|p_i-p_i^*\|_2$. Then
\begin{equation}\label{eq:KL-support}
 J(p)-J(p^*)-DJ(p^*)[p-p^*]
 \le-\frac{\Psi_{\eps,\tau}(d)}{\ln2}.
\end{equation}
Consequently the right side is at most $-\Gamma(\eps,\tau)\|p-p^*\|_2^2/\ln2$.
\end{lemma}
\begin{proof}
We separate an exact divergence identity from its uniform estimates. All divergences in the proof use natural logarithms.

\emph{Exact remainder identity:} For any positive channel $A$ and positive output reference $\nu$,
\begin{equation}\label{eq:MI-reference}
(\ln2)I(p;A)=\sum_xp(x)\KL(A_x\|\nu)-\KL(pA\|\nu).
\end{equation}
If $A_x=B_x+H_x$, where $A_x$ and $B_x$ are strictly positive
probability vectors, direct expansion of the logarithm gives
\begin{align}\label{eq:rowexpansion}
 \KL(A_x\|\nu)
 &=\KL(B_x\|\nu)+\sum_y H_x(y)\ln\frac{B_x(y)}{\nu(y)}\nonumber\\
 &\quad+\KL(A_x\|B_x).
\end{align}
For brevity put $\Lambda_i(x_i,y)=\ln(B_i(y\mid x_i)/\nu_i^*(y))$. Apply \eqref{eq:MI-reference} and \eqref{eq:rowexpansion} to \eqref{eq:Javerage}:
\begin{equation}\label{eq:Jexpanded}
\begin{split}
 (\ln2)J_i(p)
 &=\sum_{x_i}p_i(x_i)\KL(B_{i,x_i}\|\nu_i^*)\\
 &\quad+(1-\tau)\sum_{x_i,y}p_i(x_i)H_i^p(y\mid x_i)\Lambda_i(x_i,y)\\
 &\quad+(1-\tau)\sum_{x_i}p_i(x_i)\KL(C_{i,x_i}^p\|B_{i,x_i})\\
 &\quad-\tau\KL(p_iB_i\|\nu_i^*)
       -(1-\tau)\KL(p_iC_i^p\|\nu_i^*).
\end{split}
\end{equation}
Let $p(t)=p^*+t\delta$ for $0\le t\le1$, and define
\begin{equation}\label{eq:PhiDefinition}
 \Phi_i(t)=(1-\tau)\sum_{x_i,y}p_i(t)(x_i)
 H_i^{p(t)}(y\mid x_i)\Lambda_i(x_i,y).
\end{equation}
At $t=0$, $H_i^{p(0)}=0$ and
$\nu_i^*=p_i^*B_i=p_i^*C_i^{p(0)}$.
For a differentiable probability path $r(t)$ through a
strictly positive probability vector $\nu$, direct
differentiation gives
\begin{equation}\label{eq:KLderivativezero}
 \left.\frac{\dd}{\dd t}\KL(r(t)\|\nu)\right|_{t=0}
 =\sum_y r'_y(0)=0,\qquad r(0)=\nu.
\end{equation}
Indeed, differentiating
$\sum_y r_y(t)\ln(r_y(t)/\nu_y)$ at $t=0$
gives $\sum_y r'_y(0)[1+\ln(\nu_y/\nu_y)]$.
This vanishes because $\sum_y r_y(t)=1$.
Thus each of the last three terms in \eqref{eq:Jexpanded}
vanishes at $t=0$ and has zero first derivative there.
Each therefore contributes its value at $t=1$ to the
first-order Taylor remainder. The first term in
\eqref{eq:Jexpanded} is affine in $p_i(t)$ and contributes
nothing to this remainder. It remains to compute the
contribution of $\Phi_i$.
Because $H_i^{p(0)}=0$, differentiating its definition yields
\begin{equation}\label{eq:PhiDerivative}
 \begin{split}
 \Phi_i(0)&=0,\\
 \Phi_i'(0)&=(1-\tau)\sum_{x_i,y}p_i^*(x_i)
 \left.\frac{\dd}{\dd t}H_i^{p(t)}(y\mid x_i)\right|_{t=0}
 \Lambda_i(x_i,y).
 \end{split}
\end{equation}
Put
$\mathcal M(p)=\sum_i[\Phi_i(1)-\Phi_i(0)-\Phi_i'(0)]$.
Subtracting from \eqref{eq:Jexpanded} its value and derivative
at $p^*$ now gives the exact identity
\begin{align}\label{eq:exactremainder}
 &(\ln2)\{J(p)-J(p^*)-DJ(p^*)[\delta]\}\nonumber\\
 &=-\tau\sum_i\KL(p_iB_i\|\nu_i^*)
   -(1-\tau)\sum_i\KL(p_iC_i^p\|\nu_i^*)\nonumber\\
 &\quad+(1-\tau)\sum_{i,x_i}p_i(x_i)
                  \KL(C_{i,x_i}^p\|B_{i,x_i})
       +\mathcal M(p).
\end{align}
We first absorb the positive row divergences into the negative
output divergences. We then bound the mixed remainder
$\mathcal M(p)$ by differentiating the entire product input.
Finally, we lower-bound the remaining output divergences and
combine these estimates to obtain the desired global bound.


\emph{Absorbing the row divergences:} In \eqref{eq:rowcomparison}, take $P=\prod_{j\ne i}p_j$ and $Q=\prod_{j\ne i}p_j^*$. Their $G_i$ output laws are $C_{i,x_i}^p$ and $B_{i,x_i}$, respectively. Their $\mathcal B_{-i}$ output laws are product distributions, whose divergence is additive. Thus
\begin{equation}\label{eq:absorbrow}
 \KL(C_{i,x_i}^p\|B_{i,x_i})
 \le\frac{\tau^q u_i}{1-\tau}
       \sum_{j\ne i}\KL(p_jB_j\|\nu_j^*).
\end{equation}
Averaging \eqref{eq:absorbrow} over $p_i$ and summing
over $i$, using $\sum_{x_i}p_i(x_i)=1$, gives
\begin{align}\label{eq:absorbdouble}
 &(1-\tau)\sum_{i,x_i}p_i(x_i)
            \KL(C_{i,x_i}^p\|B_{i,x_i})\nonumber\\
 &\quad\le\tau^q\sum_i u_i\sum_{j\ne i}
              \KL(p_jB_j\|\nu_j^*)\nonumber\\
 &\quad=\tau^q\sum_j\ell_j\KL(p_jB_j\|\nu_j^*).
\end{align}
The last equality exchanges the finite sums and uses
$\ell_j=\sum_{i\ne j}u_i$. The coefficient on the $j$th
negative output divergence in \eqref{eq:exactremainder}
therefore changes from $-\tau$ to
$-\tau+\tau^q\ell_j=-c_j$. Hence
\begin{align}\label{eq:absorbed}
 &(\ln2)\{J(p)-J(p^*)-DJ(p^*)[\delta]\}\nonumber\\
 &\le-\sum_i c_i\KL(p_iB_i\|\nu_i^*)
      -(1-\tau)\sum_i\KL(p_iC_i^p\|\nu_i^*)\nonumber\\
 &\quad+\mathcal M(p).
\end{align}

\emph{Bounding the mixed remainder:}
Put $h_i=\|\delta_i\|_1$, so
$h_i\le\sqrt{n_i}\,d_i$ by Cauchy--Schwarz.
For $0\le t\le1$, let
$P_t(x)=\prod_{j=1}^Kp_j(t)(x_j)$, where
$x=(x_1,\ldots,x_K)$, and set
\begin{equation}\label{eq:mixedf}
 f_i(x)=\sum_y
 [W_i(y\mid x)-V_i(y\mid x_i)]\Lambda_i(x_i,y).
\end{equation}
Every entry of $B_i$ and $\nu_i^*$ lies in $[a,1]$.
Consequently,
\begin{equation}\label{eq:logratio}
 |\Lambda_i(x_i,y)|\le L:=\ln(1/a).
\end{equation}
The receiver-marginal perturbation bound says that
$\sum_y|W_i(y\mid x)-V_i(y\mid x_i)|\le2\eps$.
Therefore, term by term in \eqref{eq:mixedf},
\begin{equation}\label{eq:mixedfbound}
 |f_i(x)|\le
 \sum_y|W_i(y\mid x)-V_i(y\mid x_i)|
          |\Lambda_i(x_i,y)|
 \le2\eps L.
\end{equation}

The identity
$(1-\tau)H_i^{p(t)}=A_i^{p(t)_{-i}}-B_i$
and the averaged-channel definition
\eqref{eq:TINaveragedchannel} give
\begin{align}\label{eq:PhiProduct}
 \Phi_i(t)
 &=\sum_{x,y}P_t(x)
   [W_i(y\mid x)-B_i(y\mid x_i)]\Lambda_i(x_i,y)\nonumber\\
 &=\sum_xP_t(x)f_i(x)\nonumber\\
 &\quad-\sum_{x_i,y}p_i(t)(x_i)
   [B_i(y\mid x_i)-V_i(y\mid x_i)]
   \Lambda_i(x_i,y).
\end{align}
In the second line we inserted and subtracted
$V_i(y\mid x_i)$. The other input coordinates sum out:
\begin{equation}
 \sum_{x_{-i}}P_t(x)
 =p_i(t)(x_i)\prod_{j\ne i}\sum_{x_j}p_j(t)(x_j)
 =p_i(t)(x_i).
\end{equation}
Since $B_i,V_i,\Lambda_i$ are fixed and depend
on the input only through $x_i$, the last term in
\eqref{eq:PhiProduct} depends on $t$ only through
$p_i(t)=p_i^*+t\delta_i$. It is therefore affine
and has identically zero second derivative. Hence
\begin{equation}\label{eq:PhiSecond}
 \Phi_i''(t)=\sum_xP_t''(x)f_i(x).
\end{equation}
Each $p_j(t)=p_j^*+t\delta_j$ is affine in $t$, so its
second derivative is zero. Differentiating the full product
twice consequently gives the sum over ordered distinct pairs
\begin{equation}\label{eq:productsecond}
 P_t''(x)=
 \sum_{\substack{j,l\in[K]\\j\ne l}}
 \delta_j(x_j)\delta_l(x_l)
 \prod_{k\notin\{j,l\}}p_k(t)(x_k).
\end{equation}
There is no $j=l$ term, since $p_j''(t)=0$.
For each fixed ordered pair $(j,l)$, the other factors
are nonnegative probability distributions, and thus
\begin{align}\label{eq:productsecondnorm}
 &\sum_x\left|
 \delta_j(x_j)\delta_l(x_l)
 \prod_{k\notin\{j,l\}}p_k(t)(x_k)\right|\nonumber\\
 &\qquad=\left(\sum_{x_j}|\delta_j(x_j)|\right)
          \left(\sum_{x_l}|\delta_l(x_l)|\right)\nonumber\\
 &\qquad\quad\cdot
          \prod_{k\notin\{j,l\}}\sum_{x_k}p_k(t)(x_k)
 =h_jh_l.
\end{align}
Applying the triangle inequality to \eqref{eq:productsecond},
then using \eqref{eq:mixedfbound} and \eqref{eq:PhiSecond},
gives, uniformly for $0\le t\le1$,
\begin{equation}\label{eq:PhiSecondBound}
 |\Phi_i''(t)|
 \le2\eps L\sum_{j\ne l}h_jh_l.
\end{equation}
Finally, Taylor's formula with integral remainder yields
\begin{equation}\label{eq:PhiIntegral}
 \Phi_i(1)-\Phi_i(0)-\Phi_i'(0)
 =\int_0^1(1-t)\Phi_i''(t)\,\dd t.
\end{equation}
Using \eqref{eq:PhiSecondBound} and
$\int_0^1(1-t)\,\dd t=1/2$, each receiver contributes at
most $\eps L\sum_{j\ne l}h_jh_l$. Summing over all $K$
receivers gives the factor $K$. Moreover,
$h_j\le\sqrt{n_j}d_j$ and the definition of $M$ in
\eqref{eq:interactionmatrix} imply
$\sum_{j\ne l}h_jh_l
 \le\sum_{j\ne l}\sqrt{n_jn_l}\,d_jd_l
 =d^{\mathsf T}Md$.
Therefore
\begin{equation}\label{eq:mixedbound}
 \mathcal M(p)\le K\eps L\,d^{\mathsf T}Md.
\end{equation}

\emph{Bounding the remaining divergences from below:}
We first bound the channel variation.
Order the interfering indices increasingly and replace the
product factors one by one. The exact telescoping identity is
\begin{equation}\label{eq:producttelescoping}
 \prod_{j\ne i}p_j-\prod_{j\ne i}p_j^*
 =\sum_{j\ne i}\delta_j
  \left(\prod_{\substack{k<j\\k\ne i}}p_k\right)
  \left(\prod_{\substack{k>j\\k\ne i}}p_k^*\right).
\end{equation}
Each summand is a signed measure of total mass zero,
because $\sum_{x_j}\delta_j(x_j)=0$; its $\ell_1$ norm is
$h_j$, since all other factors are probability distributions.
The reference row $V_i(\cdot\mid x_i)$ is independent of
$x_{-i}$ and therefore cancels in each summand.
Since
$H_i^p=(A_i^{p_{-i}}-B_i)/(1-\tau)$,
the $\ell_2$ perturbation estimate in \eqref{eq:dbounds}
and the triangle inequality give, for every $x_i$,
\begin{equation}\label{eq:Htwo}
 \|H_i^p(\cdot\mid x_i)\|_2
 \le\frac{\sqrt2\eps}{1-\tau}\sum_{j\ne i}h_j.
\end{equation}
Equations~\eqref{eq:kl-lower} and \eqref{eq:singular} give
\begin{equation}\label{eq:DBlower}
 \KL(p_iB_i\|\nu_i^*)\ge\|\delta_iB_i\|_2^2\ge s_i^2d_i^2.
\end{equation}
Also
\begin{equation}\label{eq:outputdifference}
 p_iC_i^p-\nu_i^*=\delta_iB_i+p_iH_i^p.
\end{equation}
Since $p_i$ is a probability distribution, the triangle
inequality, \eqref{eq:Htwo}, and
$h_j\le\sqrt{n_j}d_j$ give
\begin{align}\label{eq:weightedHbound}
 \|p_iH_i^p\|_2
 &=\left\|\sum_{x_i}p_i(x_i)H_i^p(\cdot\mid x_i)\right\|_2\nonumber\\
 &\le\sum_{x_i}p_i(x_i)\|H_i^p(\cdot\mid x_i)\|_2\nonumber\\
 &\le\frac{\sqrt2\eps}{1-\tau}
       \sum_{j\ne i}\sqrt{n_j}d_j.
\end{align}
Apply the reverse triangle inequality to
\eqref{eq:outputdifference}, then use \eqref{eq:singular}:
\begin{align}\label{eq:outputnormlower}
 \|p_iC_i^p-\nu_i^*\|_2
 &\ge\bigl[\|\delta_iB_i\|_2-\|p_iH_i^p\|_2\bigr]_+\nonumber\\
 &\ge\left[s_i d_i-\frac{\sqrt2\eps}{1-\tau}
            \sum_{j\ne i}\sqrt{n_j}d_j\right]_+.
\end{align}
Finally, \eqref{eq:kl-lower} bounds
$\KL(p_iC_i^p\|\nu_i^*)$ below by
$\|p_iC_i^p-\nu_i^*\|_2^2$. Together with \eqref{eq:outputnormlower}, this gives
\begin{equation}\label{eq:DClower}
 \KL(p_iC_i^p\|\nu_i^*)
 \ge\left[s_i d_i-\frac{\sqrt2\eps}{1-\tau}
                  \sum_{j\ne i}\sqrt{n_j}d_j\right]_+^2.
\end{equation}
Since all $c_i\ge0$, insert \eqref{eq:DBlower}, \eqref{eq:DClower}, and \eqref{eq:mixedbound} into \eqref{eq:absorbed}. This is precisely \eqref{eq:KL-support}. Homogeneity of \eqref{eq:Psi} proves the last assertion, with the zero vector handled directly.
\end{proof}

\begin{proof}[Completion of the proof of Theorem~\ref{thm:main}]
Under \eqref{eq:tight}, Lemma~\ref{lem:tightness} and \eqref{eq:tangent} imply, for all $p\in\mathcal D$,
\begin{equation}\label{eq:globaltight}
 J(p)\le J(p^*)+DJ(p^*)[p-p^*]\le J(p^*)=F_W(p^*).
\end{equation}
Equality holds at $p=p^*$. Thus the single-letter converse and TIN lower bound coincide:
\begin{align}\label{eq:maxidentity}
 \Csum(W)&\le\max_pJ(p)=F_W(p^*)\nonumber\\
 &=\max_pF_W(p)\le\Csum(W).
\end{align}
No strict inequality in \eqref{eq:useful} or \eqref{eq:tight} was used. Finally, the mutual information chain rule gives $F_W(p)\le J(p)$. If $\Gamma>0$, Lemma~\ref{lem:tightness} makes $J(p)<J(p^*)$ whenever $p\ne p^*$, so $p^*$ is also the unique maximizer of $F_W$.
\end{proof}

\subsection{Perturbation Radius}\label{sub:radius}
\subsubsection{Existence and endpoint inclusion of the optimized certificate}
For fixed $\tau$, the useful region in
\eqref{eq:basic}--\eqref{eq:useful} is an initial interval
in $\eps$: $b$ and all $s_i$ decrease, so each $u_i$
and load $\ell_j$ increase. Throughout this region,
$c_i=\tau-\tau^q\ell_i$ is nonnegative and nonincreasing.
For fixed $d\ge0$, the quantity inside each positive part
in \eqref{eq:Psi} decreases, because $s_i d_i$ decreases
and $\eps\sum_{j\ne i}\sqrt{n_j}d_j/(1-\tau)$
increases. Its squared positive part thus decreases.
Also $a=\alpha-\eps$ decreases, so
$\eps\ln(1/a)$ increases. Since the off-diagonal
entries of $M$ and every $d_i$ are nonnegative,
$d^{\mathsf T}Md\ge0$; hence the subtracted term
$-K\eps\ln(1/a)d^{\mathsf T}Md$ is nonincreasing.
Thus $\Psi_{\eps,\tau}(d)$ is nonincreasing for each
fixed $d\ge0$. Minimizing over the fixed nonnegative
unit sphere shows that $\Gamma(\eps,\tau)$ is
nonincreasing as well. At zero perturbation,
\begin{align}\label{eq:Gamma0}
 &u_i=0,\qquad c_i=\tau,\\
 &\Psi_{0,\tau}(d)=\sum_i\sigma_i^2d_i^2,\qquad
 \Gamma(0,\tau)=\sigma^2>0.
\end{align}
Continuity therefore gives a positive feasible interval for every fixed $\tau\in(0,1)$. The explicit interval in Corollary~\ref{cor:explicit} is verified below.

The maximum in \eqref{eq:certifiedradius} is attained despite the strict inequalities in \eqref{eq:basic}. Indeed, positivity gives $\eps<\alpha$. Consider a maximizing sequence with $\eps\ge e_0>0$. The load condition implies $u_i\le1$, hence
\begin{equation}\label{eq:compactcertificate}
 N_i e_0^2\le (1-\tau)b\tau^q S_i^2.
\end{equation}
Since $0<b\le\alpha\le1$ and $1-\tau,\tau^q,S_i^2\in(0,1]$,
\eqref{eq:compactcertificate} yields, uniformly along the
maximizing sequence,
\begin{align}\label{eq:compactbounds}
 &b\ge N_i e_0^2,\qquad
 1-\tau\ge\frac{N_i e_0^2}{\alpha},\qquad
 \tau^q\ge\frac{N_i e_0^2}{\alpha},\\
 &S_i^2\ge\frac{N_i e_0^2}{\alpha}\qquad(i\in[K]).
\end{align}
In particular, $\tau$ and $1-\tau$ cannot approach zero,
and $b$ remains positive. For each $j$ choose some $i\ne j$.
Since $S_i$ contains $s_j$ and all its other factors are at
most one, $s_j\ge S_i$; the last inequality above bounds
$s_j$ away from zero. Also $0\le\eps<\alpha$ along the sequence.
Thus a convergent subsequence has its limit inside the
strict domain \eqref{eq:basic}, not on its boundary.
The loads are continuous there. Since $\Psi_{\eps,\tau}(d)$
is jointly continuous and the nonnegative unit sphere is
compact and fixed, its minimum $\Gamma(\eps,\tau)$ is
continuous as well. The limit therefore satisfies
\eqref{eq:useful}--\eqref{eq:tight}, proving that the
maximum is attained and the certified closed ball includes
its endpoint.

\subsubsection{Proof of the closed-form corollary}
Set $\tau=1/2$ and $s=\sigma-\sqrt{2n}\eps$. If $0\le\eps\le\eps_{\mathrm{cf}}$, then
\begin{equation}\label{eq:freeze}
 a\ge\frac{3\alpha}4,\qquad b=\alpha-3\eps\ge b_0>0,
 \qquad \ln(1/a)\le L_0.
\end{equation}
The denominator in \eqref{eq:radiusU} exceeds
$q\sigma^{q-1}\sqrt{2n}$, hence
$\eps_{\mathrm U}<\sigma/(q\sqrt{2n})
 \le\sigma/\sqrt{2n}$ and all $s_i\ge s>0$.
The definition of $\eps_{\mathrm U}$ also gives the
rearranged inequality
\begin{equation}\label{eq:radiusUrearranged}
 \sigma^q-q\sigma^{q-1}\sqrt{2n}\eps
 \ge\eps\sqrt{\frac{2^{q+1}qn^q}{b_0}}.
\end{equation}
Apply Bernoulli's inequality $(1-x)^q\ge1-qx$
with $x=\sqrt{2n}\eps/\sigma\in[0,1]$
and then \eqref{eq:radiusUrearranged} to obtain
\begin{equation}\label{eq:Bernoulliuse}
 s^q\ge\sigma^q-q\sigma^{q-1}\sqrt{2n}\eps
 \ge\eps\sqrt{\frac{2^{q+1}q n^q}{b_0}}.
\end{equation}
Since $N_i\le n^q$ and $S_i\ge s^q$,
\eqref{eq:Bernoulliuse} implies
$s^{2q}\ge2^{q+1}qn^q\eps^2/b_0$.
It follows that
$2^{q+1}n^q\eps^2/(bs^{2q})
 \le b_0/(qb)\le1/q$, where $b\ge b_0$.
In particular, the individual loads satisfy
\begin{equation}\label{eq:uniformloads}
 u_i\le\frac{2^{q+1}n^q\eps^2}{b s^{2q}}\le\frac1q,
 \qquad \ell_j\le1.
\end{equation}
At $\eps=0$ these assertions hold directly. Thus the useful-genie condition is verified.

We next bound \eqref{eq:Gamma} without retaining its nonnegative $c_i$ terms. Let $A$ be the $K\times K$ matrix with $A_{ij}=\sqrt{n_j}$ for $i\ne j$ and $A_{ii}=0$. Write $A=(\mathbf1\mathbf1^{\mathsf T}-I)
\operatorname{diag}(\sqrt{n_j})$.
The matrix $\mathbf1\mathbf1^{\mathsf T}-I$ has
eigenvalue $q=K-1$ in the all-ones direction and
eigenvalue $-1$ on its orthogonal complement.
Its operator norm is therefore $q$, while
$\|\operatorname{diag}(\sqrt{n_j})\|_{\mathrm{op}}
 \le\sqrt n$. Hence
\begin{equation}\label{eq:Anorm}
 \|A\|_{\mathrm{op}}\le q\sqrt n.
\end{equation}
For $d\ge0$ of unit norm, let $z_i=[s_i d_i-2\sqrt2\eps(Ad)_i]_+$.
Since $[x]_+\ge x$, $s_i\ge s$, and $d_i\ge0$, we obtain
the coordinatewise inequality $sd\le z+2\sqrt2\eps Ad$.
Taking Euclidean norms, using norm monotonicity on
the nonnegative orthant, and applying the triangle
inequality yields
$\|z\|_2\ge s\|d\|_2-2\sqrt2\eps\|Ad\|_2
 \ge s-2q\sqrt{2n}\eps$.
Therefore
\begin{equation}\label{eq:uniformz}
 \|z\|_2\ge s-2q\sqrt{2n}\eps=\sigma-C_K\eps.
\end{equation}
The right side is positive because \eqref{eq:radiusG} gives $\eps_{\mathrm G}<\sigma/(2C_K)$. Using $d^{\mathsf T}Md\le\rho$ and \eqref{eq:freeze},
\begin{align}\label{eq:uniformGamma}
 \Gamma(\eps,1/2)
 &\ge\tfrac12(\sigma-C_K\eps)^2-K\rho L_0\eps\nonumber\\
 &\ge\tfrac12\sigma^2-(\sigma C_K+K\rho L_0)\eps\ge0.
\end{align}
To see the last steps explicitly, expand
$\frac12(\sigma-C_K\eps)^2
 =\frac12\sigma^2-\sigma C_K\eps+\frac12C_K^2\eps^2$
and discard the final nonnegative term. The definition
\eqref{eq:radiusG} gives
$(\sigma C_K+K\rho L_0)\eps\le\sigma^2/2$.
Thus \eqref{eq:uniformGamma} is nonnegative for
all $0\le\eps\le\eps_{\mathrm{cf}}$, including the endpoint.
This proves Corollary~\ref{cor:explicit}.


\subsubsection{A scalar form for equal link parameters}
When all $n_i=n$ and all $\sigma_i=\sigma$, the minimization in \eqref{eq:Gamma} has an exact scalar evaluation. Define
\begin{align}
&s=\sigma-\sqrt{2n}\eps,\qquad
 u=\frac{n^q\eps^2}{(1-\tau)b\tau^q s^{2q}},\label{eq:equalu}\\
 &c=\tau-\tau^q q u,\qquad
 \lambda=\left[s-\frac{q\sqrt{2n}\eps}{1-\tau}\right]_+.
 \label{eq:equallambda}
\end{align}
For $0<\tau<1$, $b>0$, $s>0$, and $qu\le1$,
\begin{equation}\label{eq:equalGamma}
 \Gamma(\eps,\tau)=cs^2+(1-\tau)\lambda^2
                   -Knq\eps\ln\frac1{\alpha-\eps}.
\end{equation}
For an arbitrary nonnegative unit vector $d$, the same
norm argument as \eqref{eq:uniformz} gives
$\|z\|_2\ge[s-q\sqrt{2n}\eps/(1-\tau)]_+=\lambda$
for the positive-part vector $z$ in \eqref{eq:Psi}.
Also $d^{\mathsf T}Md\le\lambda_{\max}(M)=nq$
and $\sum_i cs^2d_i^2=cs^2$.
Together these prove the lower bound in \eqref{eq:equalGamma}.
To verify that no loss occurs, choose the uniform direction
\begin{align}\label{eq:equalattainment}
 &d_i=1/\sqrt K,\qquad d^{\mathsf T}Md=nq,\\
 &\left[s d_i-\frac{\sqrt2\eps}{1-\tau}
      \sum_{j\ne i}\sqrt n\,d_j\right]_+
      =\lambda/\sqrt K,\\
 &\sum_i cs^2d_i^2=cs^2.
\end{align}
Then $\|z\|_2^2=\lambda^2$ and the interaction quadratic
form equals $nq$, so every bound is an equality simultaneously.
Substitution in \eqref{eq:Psi} proves the exact formula,
including the case $\lambda=0$. The reference channels need not be identical; equality of the stated dimension and singular-value parameters suffices.


\section{Conclusion}\label{sec:conclusion}
We have established exact TIN sum-capacity optimality throughout a neighborhood of positive radius around every regular finite-alphabet product channel with a fixed number of users. The resulting sufficient conditions are explicit and readily verifiable from the channel transition probabilities. They identify a physically interpretable class of noisy-interference channels arising from weak interactions among otherwise independent communication links, thereby connecting the abstract information-theoretic notion of noisy interference with channel models that arise naturally in practice. We have not attempted to characterize the largest neighborhood in which TIN remains sum-capacity optimal. Moreover, the strict-positivity assumption can be relaxed by defining neighborhoods using receiver-marginal KL divergence from the perturbed channel to the reference. Finiteness of this divergence preserves the zero-valued reference transition probabilities, allowing the analysis to extend to product channels whose components have full row rank but may contain zeros, such as nondegenerate erasure channels and Z-channels.

More broadly, stability analysis offers a systematic approach to exploring less understood channel models through perturbations of models with known capacity characterizations. This perspective may also be fruitful for other families of interference channels, including those in the strong-interference regime, and for other network models, such as broadcast channels. By examining whether achievable rates and matching converse bounds remain aligned under perturbations, it brings techniques from optimization, including curvature estimates and sensitivity analysis, into the study of information-theoretic optimality. Such an approach could reveal new capacity classes and open further avenues for research in network information theory.

\appendices

\section{Proof of Lemma \ref{lem:farsani}}\label{app:lemma1}

For a sequence $A^N$, define its total correlation \cite{Watanabe60}, in bits, by
\begin{equation}\label{eq:TC}
 \TC(A^N)=\sum_{t=1}^N H(A_t)-H(A^N).
\end{equation}
For any conditioning variable $D$, define
\begin{equation}\label{eq:TCconditionaldefinition}
 \TC(A^N\mid D)=\sum_{t=1}^N H(A_t\mid D)-H(A^N\mid D).
\end{equation}
We first prove two comparisons used only in this converse.

First, suppose at each time $t$ a channel $A_t\to B_t$ is less noisy than $A_t\to C_t$, and the channels are used memorylessly over time. For any input-sequence law, possibly temporally dependent,
\begin{equation}\label{eq:TCorder}
 \TC(B^N)\ge\TC(C^N).
\end{equation}
To prove this, choose a joint coupling of $B_t,C_t$ with their
given output marginals, independently across time given $A^N$.
For $k=0,\ldots,N$, define
$Z^{(k)}=(B_1,\ldots,B_k,C_{k+1},\ldots,C_N)$.
For $k=1,\ldots,N$, let
$U_k=(B_1,\ldots,B_{k-1},C_{k+1},\ldots,C_N)$.
Conditional memorylessness yields $U_k-A_k-(B_k,C_k)$ even for
temporally correlated inputs: once $A_k$ is known, the randomness
at time $k$ is independent of all the remaining channel outputs.
The less-noisy condition thus applies with auxiliary $U_k$.
Expanding total correlation and canceling the terms containing
only $U_k$ gives
\begin{align}\label{eq:TCreplace}
 \TC(Z^{(k)})-\TC(Z^{(k-1)})
 &=\TC(U_k,B_k)-\TC(U_k,C_k)\nonumber\\
 &=I(U_k;B_k)-I(U_k;C_k)\ge0.
\end{align}
Summing over $k$ proves \eqref{eq:TCorder}. Nothing requires
the single-letter channels to be identical across time, so
a fixed state sequence may index the channels.

Second, let $\mathcal E_\eta(A^N)$ erase each coordinate
independently with retention probability $\eta$, and let
$S\subseteq[N]$ be the retained set. Denote binary entropy in
bits by $h_2$. The erasure symbol identifies whether a
coordinate was retained, and its retention indicator is
independent of $A_t$. Hence
\begin{equation}
 H(\mathcal E_\eta(A_t))=h_2(\eta)+\eta H(A_t).
\end{equation}
The entire erased sequence likewise determines $S$.
Conditioned on $S$, the unerased entries have the law of
$A_S$, since the pattern is independent of $A^N$. Moreover,
the $N$ retention indicators are independent, so
\begin{equation}
 H(\mathcal E_\eta(A^N))
 =H(S)+H(A_S\mid S)
 =Nh_2(\eta)+\mathbb E H(A_S),
\end{equation}
where the expectation is taken with respect to the random retained
set $S$.
Subtracting the block entropy from the sum of the coordinate
entropies cancels the two $Nh_2(\eta)$ terms and yields
\begin{equation}\label{eq:TCerasureexact}
 \TC(\mathcal E_\eta(A^N))
 =\eta\sum_tH(A_t)-\mathbb E H(A_S).
\end{equation}
Order the coordinates increasingly. Conditioning reduces entropy, so
\begin{align}\label{eq:subsetentropy}
 \mathbb E H(A_S)
 &=\mathbb E\sum_{t\in S}H(A_t\mid A_{S\cap[t-1]})\nonumber\\
 &\ge\mathbb E\sum_{t\in S}H(A_t\mid A^{t-1})
 =\eta H(A^N).
\end{align}
Thus
\begin{equation}\label{eq:TCerasure}
 \TC(\mathcal E_\eta(A^N))\le\eta\TC(A^N).
\end{equation}

Now take an arbitrary reliable length-$N$ code and fix a receiver $i$.
Independent messages and separate encoders make the entire input sequences
independent across users. The memoryless product virtual channel
\eqref{eq:virtualproduct} consequently makes the sequences $T_j^N$
independent across users and $T_{-i}^N$ independent of $X_i^N$.
In particular, for every $x_i^N$ of positive probability,
the conditional law of $X_{-i}^N$ given $X_i^N=x_i^N$ equals its
unconditional law; it may still have arbitrary dependence across time.

Fix such an $x_i^N$. At time $t$, treat $x_{i,t}$ as a fixed
state and take $X_{-i,t}$ as the common input. Let
$\widetilde T_{i,t}$ and $Z_{i,t}$ denote the outputs of
$E_i$ and $L_{i,x_{i,t}}$, respectively.
By \eqref{eq:farsani-conditions}, the former channel is
less noisy than the latter for every choice of $x_{i,t}$.
Conditional on $X_i^N=x_i^N$, these channels are used
memorylessly over time, even though $X_{-i}^N$ need not
be time-independent. Thus \eqref{eq:TCorder} applies to
this conditional input law.

Write $\widetilde T_i^N=(\widetilde T_{i,t})_{t=1}^N$.
By the definition of $E_i$, this sequence is obtained by
erasing each time-coordinate of the vector sequence
$T_{-i}^N=(T_{-i,t})_{t=1}^N$ independently with retention
probability $\eta_i$. The erasure decisions are independent
of the inputs and virtual-channel randomness.
Applying \eqref{eq:TCerasure} to the same conditional law gives
\begin{align}\label{eq:conditionalTCpointwise}
 \TC(Z_i^N\mid X_i^N=x_i^N)
 &\le\TC(\widetilde T_i^N\mid X_i^N=x_i^N)\nonumber\\
 &\le\eta_i\TC(T_{-i}^N\mid X_i^N=x_i^N)\nonumber\\
 &=\eta_i\TC(T_{-i}^N).
\end{align}
The last equality follows from $T_{-i}^N\perp X_i^N$.


The virtual sequences $T_j^N$, $j\ne i$, are mutually independent
as entire sequences, not necessarily within each sequence.
Consequently, their single-time entropies and their block entropy
both add across users:
\begin{align}\label{eq:TCvirtualsum}
 \TC(T_{-i}^N)
 &=\sum_tH(T_{-i,t})-H(T_{-i}^N)\nonumber\\
 &=\sum_{j\ne i}\left[\sum_tH(T_{j,t})-H(T_j^N)\right]\nonumber\\
 &=\sum_{j\ne i}\TC(T_j^N).
\end{align}
Average \eqref{eq:conditionalTCpointwise} over $X_i^N$, using the
definition of conditional total correlation, and substitute
\eqref{eq:TCvirtualsum}. We obtain
\begin{equation}\label{eq:conditionalTC}
 \TC(Z_i^N\mid X_i^N)
 \le\eta_i\TC((T_{-i,t})_{t=1}^N)
 =\eta_i\sum_{j\ne i}\TC(T_j^N).
\end{equation}
Total correlation admits the divergence representation
\begin{equation}\label{eq:TCrelativeentropy}
 \TC(Z_i^N)=\KL\left(P_{Z_i^N}\Big\|
                  \prod_t P_{Z_{i,t}}\right)/\ln2.
\end{equation}
The coordinatewise projection $Z_{i,t}\mapsto T_{i,t}$
maps both arguments of the divergence to the corresponding
joint and product laws of $T_i^N$. Data processing therefore gives
\begin{equation}\label{eq:projectionTC}
 \TC(Z_i^N)\ge\TC(T_i^N).
\end{equation}
Moreover, independence of the users' input sequences and channel
memorylessness imply
$H(Z_{i,t}\mid X_i^N)=H(Z_{i,t}\mid X_{i,t})$:
the other inputs are independent of $X_i^N$, and the time-$t$
output depends only on the current inputs.
Insert the definitions of unconditional and conditional total
correlation into the block mutual information to obtain
\begin{align}\label{eq:blockidentity}
 I(X_i^N;Z_i^N)
 &=H(Z_i^N)-H(Z_i^N\mid X_i^N)\nonumber\\
 &=\sum_t[H(Z_{i,t})-H(Z_{i,t}\mid X_{i,t})]\nonumber\\
 &\quad-\TC(Z_i^N)+\TC(Z_i^N\mid X_i^N)\nonumber\\
 &=\sum_t I(X_{i,t};Z_{i,t})-\TC(Z_i^N)\nonumber\\
 &\quad+\TC(Z_i^N\mid X_i^N).
\end{align}
Summing over $i$ and using \eqref{eq:conditionalTC} and
\eqref{eq:projectionTC} bounds the entire cross-time remainder by
\begin{align}\label{eq:TCbudgetcancellation}
 &\sum_i[-\TC(Z_i^N)+\TC(Z_i^N\mid X_i^N)]\nonumber\\
 &\ \le-\sum_j\TC(T_j^N)
      +\sum_i\eta_i\sum_{j\ne i}\TC(T_j^N)\nonumber\\
 &\ =\sum_j\left(-1+\sum_{i\ne j}\eta_i\right)
        \TC(T_j^N)\le0.
\end{align}
The last step uses \eqref{eq:budget} and nonnegativity of
total correlation. We conclude that
\begin{equation}\label{eq:singlelettersum}
 \sum_i I(X_i^N;Z_i^N)\le\sum_t\sum_iI(X_{i,t};Z_{i,t}).
\end{equation}
For completeness, Fano's inequality \cite{Fano61}
and data processing for the message--codeword--output
Markov chain give, for each receiver,
\begin{align}\label{eq:Fanoexplicit}
 &(1-P_{e,i}^{(N)})\log_2M_{i,N}\nonumber\\
 &\quad\le I(M_i;Y_i^N)+1
       \le I(X_i^N;Z_i^N)+1.
\end{align}
Since the output alphabets and $K$ are fixed,
these inequalities keep the normalized message logarithms
bounded as the errors vanish. Dividing by $N$, summing
over $i$, and letting the error probabilities tend to zero
gives a sum-rate upper bound of
$N^{-1}\sum_i I(X_i^N;Z_i^N)+o(1)$.
For each time $t$, independent user messages and separate
encoders make the joint input law a product distribution.
Writing
$p^{(t)}=(P_{X_{1,t}},\ldots,P_{X_{K,t}})\in\mathcal D$
for the corresponding tuple of input marginals, we obtain
$\sum_i I(X_{i,t};Z_{i,t})
 =J(p^{(t)})\le\max_{p\in\mathcal D}J(p)$.
Combining this with \eqref{eq:singlelettersum} and taking
$N\to\infty$ proves \eqref{eq:geniebound}.

\section*{Acknowledgment}
This work was supported in part by a grant of access to OpenAI models
through the ChatGPT for Academic Researchers program. GPT-6 Astra
was used to assist with the perturbation analysis in the proof of
Theorem~\ref{thm:main} and with manuscript editing. The authors
are responsible for the problem formulation, the overall proof
strategy, and the correctness of the results.

\end{document}